\documentclass{WileyMSP-template}
\usepackage{amsmath}
\begin{document}

\pagestyle{fancy}
\rhead{\includegraphics[width=2.5cm]{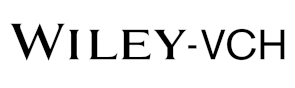}}

\title{Ultrafast nonlinear dynamics of indium tin oxide nanocrystals \\
probed via fieldoscopy}

\maketitle


\author{Andreas Herbst}
\author{Anchit Srivastava}
\author{Kilian Scheffter}
\author{Soyeon Jun}
\author{Steffen Gommel}
\author{Luca Rebecchi}
\author{Sidharth Kuriyil}
\author{Andrea Rubino}
\author{Nicolò Petrini}
\author{Ilka Kriegel}
\author{Hanieh Fattahi}



\begin{affiliations}
A. Herbst, A. Srivastava, K. Scheffter, S. Jun, S. Gommel, Dr. H. Fattahi\\
Max Planck Institute for the Science of Light, Staudtstraße 2, 91058 Erlangen, Germany\\
Friedrich-Alexander-Universität Erlangen-Nürnberg, Staudtstraße 7, 91058 Erlangen, Germany\\
K. Scheffter, S. Jun, S. Gommel\\
Erlangen Graduate School in Advanced Optical Technologies (SAOT), 91052 Erlangen, Germany\\
Dr. N. Petrini, Dr. Luca Rebecchi, S. Kuriyil, Dr. A. Rubino, Prof. I. Kriegel\\
Department of Applied Science and Technology (DISAT), Politecnico di Torino, Corso Duca degli Abruzzi, 24, 10129, Torino, Italy\\
S. Kuriyil, Dr. A. Rubino, Prof. I. Kriegel\\
Funtional Nanosystems, Istituto Italiano di Tecnologia, via Morego, 30
16163 Genova, Italy\\
Email Address:
hanieh.fattahi@mpl.mpg.de\\
\end{affiliations}

\keywords{Fieldoscopy, near-petahertz field sampling, indium tin oxide, metal oxide nano crystal, localized surface plasmon resonance, plasmonic semiconductor nano crystals, optical switch}

\begin{abstract}
Scalable, high-speed, small-footprint photonic switching platforms are essential for advancing optical communication. An effective optical switch must operate at high duty cycles with fast recovery times, while maintaining substantial modulation depth and full reversibility. Colloidal nanocrystals, such as indium tin oxide (ITO), offer a scalable platform to meet these requirements. In this work, the transmission of ITO nanocrystals near their epsilon-near-zero wavelength is modulated by two-cycle optical pulses at a repetition rate of one megahertz. The modulator exhibits a broad bandwidth spanning from 2\,$\mu$m to 2.5\,$\mu$m. Sensitive fieldoscopy measurements resolve the transient electric-field response of the ITO for the first time, showing that the modulation remains reversible for excitation fluences up to 1.2\,$\frac{\text{mJ}}{\text{cm}^2}$ with a modulation depth of 10\%, and becomes fully irreversible beyond 3.3\,$\frac{\text{mJ}}{\text{cm}^2}$, while reaching modulation depth of up to 20\%. Field sampling further indicates that at higher excitation fluences, the relative contribution from the first cycle of the optical pulses is reduced. These findings are crucial for the development of all-optical switching, telecommunications, and sensing technologies capable of operating at terahertz switching frequencies.
\end{abstract}

\section{Introduction}

Scalable, high-speed photonic switching platforms are essential for the advancement of next-generation optical communication technologies \cite{kinsey2015epsilon, sennary2025light, sweatlock2012vanadium, guo2016ultrafast, sommer2016attosecond, hui2023ultrafast,hui2022attosecond, alam2016large, kim2012flexible, abb2014hotspot, ni2016ultrafast}. For applications operating at telecommunication wavelengths, materials that support plasmonic resonances in the short-wavelength infrared (SWIR) range are particularly advantageous. Nanocrystals composed of conducting metal oxides can enable large-amplitude all-optical switching within this spectral region. This capability, based on ultrafast modulation of transmittance or reflectance, has demonstrated significant promise across a variety of near-to-far-infrared active material systems. Furthermore, the compatibility of these nanocrystals with solution-based processing and scalable fabrication methods enhances their appeal as a versatile and robust platform for high-performance, ultrafast photonic switching devices \cite{kinsey2015epsilon, sweatlock2012vanadium,choi2016exploiting, kim2011full,Fafarman2013}.

Heavily doped conducting metal oxides, such as indium tin oxide (ITO), exhibit distinctive electronic and optical characteristics due to their ability to sustain high carrier concentrations, sufficient to populate the conduction band even at room temperature \cite{kriegel_plasmonic_2017}. In nanocrystal form, these materials support broadband plasmonic absorption at the localized surface plasmon resonance (LSPR) frequency, while preserving optical transparency across a wide spectral window between the LSPR and the fundamental band gap, behavior reminiscent of metallic nanoparticles \cite{halas_plasmons_2011}. The optical response of ITO nanocrystals is primarily dictated by a wide band gap in the ultraviolet and an LSPR located in the near-to mid-infrared range. A key advantage of ITO nanocrystals lies in the remarkable tunability of their LSPR, which can be precisely modulated through control of doping concentration, and the dielectric environment \cite{garcia_dynamically_2011}. This dynamic control contrasts with conventional metallic nanostructures, whose plasmonic resonances are largely constrained by intrinsic material properties and geometric parameters \cite{scotognella_plasmonics_2013, luther_localized_2011}.

The LSPR in ITO nanocrystals arises from the collective oscillation of carriers in the conduction band, driven by a transient electric field. Upon optical excitation, these oscillations lead to localized heating of the conduction band electrons, which in turn induces a transient modulation of the material’s optical properties. The resulting relaxation dynamics proceed in two distinct stages: an ultrafast regime dominated by carrier–carrier scattering occurring on sub-picosecond timescales, followed by a slower regime driven by carrier–phonon interactions, typically spanning hundreds of picoseconds \cite{scotognella_plasmonics_2013}. In addition to LSPR behavior, ITO exhibits an epsilon-near-zero (ENZ) wavelength in SWIR  \cite{wu_epsilon-near-zero_2021, issah_epsilon-near-zero_2023, kim_hierarchically_2023,bohn2021all}. ITO nanocrystals synthesized via colloidal chemistry offer a scalable and versatile source of optically active materials. Their dispersibility in common solvents enables straightforward integration into devices through solution-based techniques such as spray coating or spin coating. This processability allows for the fabrication of patternable, large-area optoelectronic components on a wide variety of substrates, including flexible and transparent platforms \cite{ choi2016exploiting, kim2011full,Rebecchi2025}. 

Excitation of the LSPR in ITO nanocrystals has been shown to dynamically modulate the material’s dielectric function at optical frequencies between the bandgap and the LSPR. This modulation results in increased reflectivity and absorption, enabling active control over the material’s optical response \cite{Guo2017SolutionProcessed,guo2016ultrafast, blemker_modulation_2020, tice_ultrafast_2014, scotognella_plasmonics_2013}. A similar behavior is observed in indium-doped cadmium oxide nanocrystals, where LSPR excitation at the ENZ wavelength induces a transient redshift of the plasmon resonance. This shift leads to enhanced absorption for frequencies below the LSPR ($\omega < \omega_{\mathrm{LSPR}}$) and bleaching for frequencies above it ($\omega > \omega_{\mathrm{LSPR}}$) \cite{diroll_large_2016}.

An effective optical switch must operate at high duty cycles with fast recovery time, while maintaining a substantial modulation depth and full reversibility. However, prior studies have largely been restricted to low-duty-cycle excitation pulses in the kilohertz range. Although these investigations demonstrated high modulation depths and achieved significant optical switching up to a certain threshold, they were limited in repetition rate, constraining their relevance for practical high-speed applications. Furthermore, pump–probe spectroscopy has been employed to investigate the buildup and relaxation dynamics of LSPR in nanocrystals and nanostructures, particularly within the SWIR spectral range, since field-sampling methods have previously been restricted to the mid-infrared and terahertz spectral ranges \cite{Fischer:21, scheffter2024compressed}. As a result, insight into ultrafast dynamics on sub-cycle timescales has remained limited. Recent progress in field-sampling metrology has expanded the accessible detection bandwidth, now extending into the petahertz regime \cite{herbst_recent_2022, srivastava_near-petahertz_2024,Cho2019,Park:18,Sederberg2020,Zimin:21,Liu:21,Liu2021,Beckh2021,Alismail:20,zimin_ultra-broadband_2022,Bionta2021,Korobenko:20,Kowligy:19,Riek2017,Saito:18,keiber_electro-optic_2016,Liu:24,Kubullek:2025}. These advances have enabled direct, time-resolved measurements of plasmonic responses in metallic nanoparticles and nanostructures at optical frequencies \cite{wong_far-field_2024, zimin2023electric, mamaikin2022electric}.

This work investigates the ultrafast nonlinear plasmonic response of ITO nanocrystals under intense, ultrashort excitation near their ENZ wavelength and at megahertz repetition rates. Using femtosecond Fieldoscopy \cite{srivastava_near-petahertz_2024,srivastava_180_2024}, we resolve sub-cycle transients with attosecond precision, directly tracking the field dynamics. The results show fluence-dependent modulation with a reversible 10\% depth at megahertz duty cycles, fulfilling the key requirements for effective high-speed optical switching.

\section{Results and discussion}

\begin{figure}[htb]
    \centering
    \includegraphics{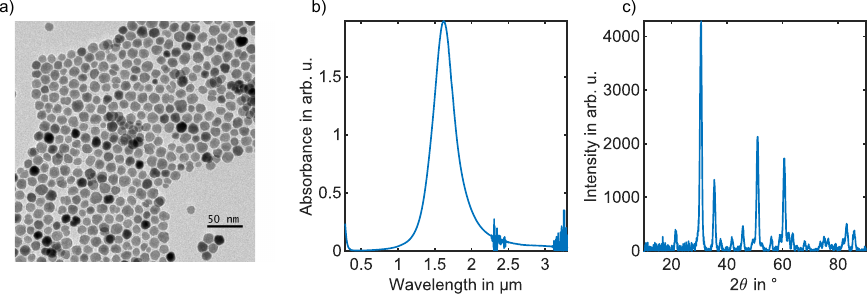}
    \caption{SI. a) Representative TEM image of the ITO nanocrystals batch employed for film deposition. b) Absorbance spectrum of 10\,µL of a 1\,$\frac{\text{mg}}{\text{mL}}$ dispersion diluted in 700\,µL of hexane. c) XRD pattern of the resulting ITO film, displaying the characteristic diffraction peaks of crystalline ITO \cite{Rebecchi2025}.}
    \label{SI1}
\end{figure}

The samples were prepared by depositing a multilayer of ITO nanocrystals via the dip coating technique onto 150\,µm-thick borosilicate Menzel glass substrates, followed by annealing at 100\,°C (see Methods and Figure \ref{SI1} SI). A scanning electron microscope (SEM) image of the resulting ITO-coated sample is shown in \textbf{Figure \ref{fig:1}a}, showing nanocrystals with an average size of 14\,nm (\textbf{Figure \ref{fig:1}b}). To characterize the transient electric field of ITO nanocrystals near their ENZ wavelength, the samples were excited using carrier-envelope phase (CEP)-stable pulses with a duration of 10.7\,fs and a spectral range spanning from 120\,THz to 200\,THz, corresponding to the wavelength range of 1.5\,µm to 2.5\,µm. The spectrum of the excitation pulses, the LSPR spectrum of the ITO nanocrystals and a schematic of the experimental setup are shown in \textbf{Figure \ref{fig:1}c} and \textbf{Figure \ref{fig:1}d}. 

\begin{figure}[t]
    \centering
    \includegraphics[width=0.85\textwidth]{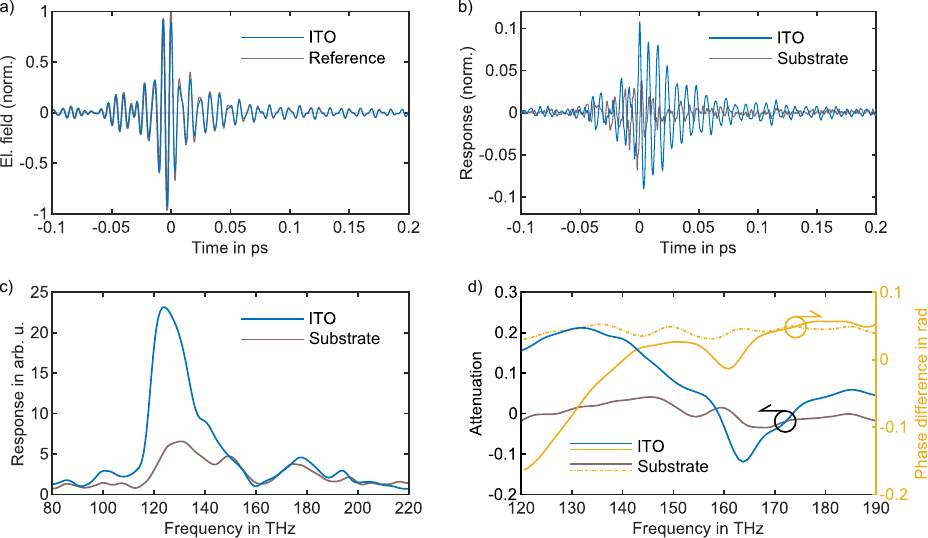}
    \caption{a) Electric field of the transmitted excitation pulses from the ITO and substrate sample at the fluence of $75\,\frac{\mu\text{J}}{\text{cm}^2}$. The substrate measurements serve as reference measurements in this study. b) Calculated sample response obtained by subtracting the resolved electric field of the ITO from the reference at the same fluence (blue curve). Prior to subtraction, the two fields were temporally aligned by maximizing their cross-correlation within the window of –80\,fs to 10\,fs. The gray curve represents the averaged residual electric field from four consecutive reference measurements. c) Fourier transform of the ITO response, indicating the bandwidth of the response. The amplitude shows the magnitude of the attenuation in the ITO. d) The complex transmission spectrum of the sample, which is derived from the Fourier transformation of the transmitted electric field of ITO and the reference separately. The yellow curve indicates the spectral phase.}
    \label{fig:ITO vs Glass}
\end{figure}

To investigate the nonlinear dynamics of the ITO nanocrystals, the excitation pulses were focused onto the sample at varying pump fluences, and the electric field of the transient far-field response was resolved. Fieldoscopy (see Method) was used to access the electric field of the transmitted excitation pulses from the sample directly, with a temporal resolution of 90\,attoseconds. \textbf{Figure \ref{fig:ITO vs Glass}a} shows the field-resolved transient of the ITO ($E_S$) and the substrate ($E_R$) at the fluence of $75\,\frac{\mu\text{J}}{\text{cm}^2}$. Reference measurements on the bare substrate were performed to isolate the sample response from any linear or nonlinear contributions of the borosilicate and the spectral response of the filters. The blue curve in \textbf{Figure \ref{fig:ITO vs Glass}b} shows the sample response obtained by subtracting the transmitted ITO electric field from the reference electric field, normalized to the peak of the reference transient electric field ($E_{Response}(t) = \frac{E_R(t)-E_S(t)}{\max(|E_R(t)|)}$). The resolved electric field of the response provides information equivalent to a virtual interferometric measurement, capturing both attenuation and dephasing relative to the excitation pulse, while also revealing the timescales over which these changes occur. Prior to subtraction, the two electric fields were temporally overlapped by maximizing their cross-correlation within the window from -60\,fs to -10\,fs. The gray curve represents the residual electric field, obtained by averaging and subtracting four consecutive substrate measurements, which reflects detection noise arising from source fluctuations and temporal jitter. \textbf{Figure \ref{fig:ITO vs Glass}c} shows the Fourier transformation of the response and the residual field for the temporal window of -350\,fs to 750\,fs. The spectral amplitude of the response exhibits attenuation in the 120\,THz to 150\,THz spectral range, corresponding to the LSPR of the ITO. Alternatively, the amplitude of the transmission spectrum of the sample can be obtained by Fourier transforming each transient electric field and evaluating the relative spectral response as $\frac{\Delta I(\omega)}{I_R(\omega)}=\frac{I_R(\omega) - I_S(\omega)}{I_R(\omega)}$, as shown in \textbf{Figure \ref{fig:ITO vs Glass}d}. The retrieved spectral phase is obtained by calculating the argument of $\frac{E_{\omega}(S)}{E_{\omega}(R)}$ and indicates the induced spectral dispersion to the excitation pulses by the sample and the LSPR absorption.  

\begin{figure}[t]
    \centering
    \includegraphics{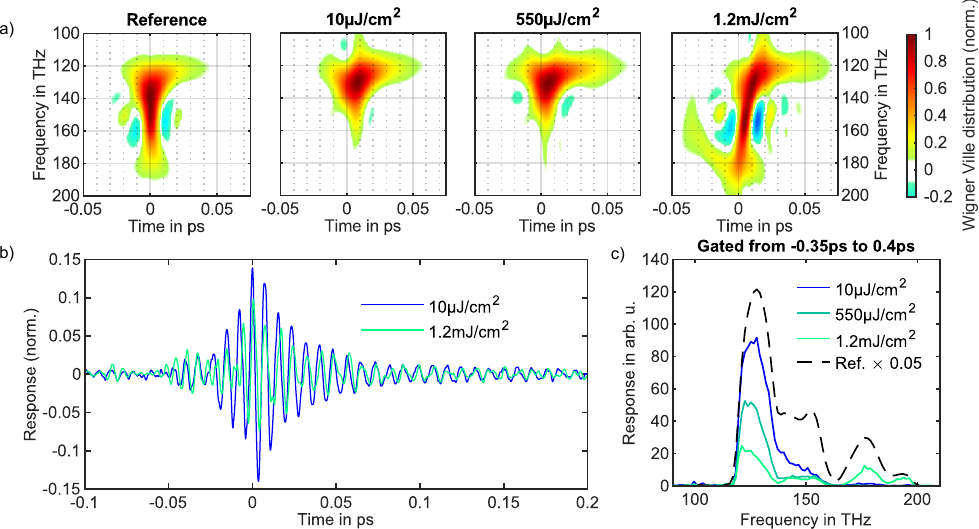}
\caption{a) Normalized Wigner–Ville distribution for the substrate (reference) and the sample responses at fluences between $10\,\frac{\mu\text{J}}{\text{cm}^2}$ and $1.2\,\frac{\text{mJ}}{\text{cm}^2}$. b) Temporal response of the sample at $10\,\frac{\mu\text{J}}{\text{cm}^2}$ and $1.2\,\frac{\text{mJ}}{\text{cm}^2}$ fluences. c) Spectral response of the sample at fluences between $10\,\frac{\mu\text{J}}{\text{cm}^2}$ and $1.2\,\frac{\text{mJ}}{\text{cm}^2}$.}
    \label{fig:3}
\end{figure}

\begin{figure}[htb]
    \centering
    \includegraphics{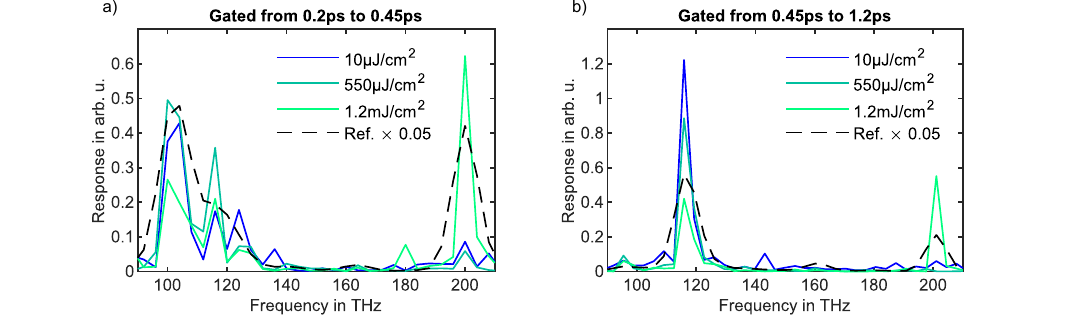}
    \caption{SI. Spectral response of the sample calculated within the time windows a) $0.2$\,ps to $0.45\,$ps and b) $0.45\,$ps to $1.2\,$ps. The dashed line represents the spectrum of the reference field. Resonances of atmospheric water molecules appear at 120\,THz and 160\,THz, while the absorption at 200\,THz originates from the dielectric beam splitter in the beam path. The response of the nanocrystals at different fluences is imprinted on these long-lived spectral fingerprints of atmospheric molecules and can be exploited to probe the transient response of the sample on longer time scales.}
    \label{SI2}
\end{figure}

\begin{figure}[htb]
    \centering
    \includegraphics{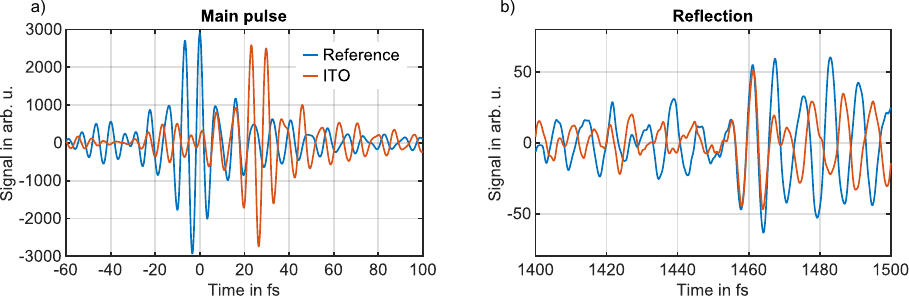}
    \caption{SI. The reflection of the excitation pulse at the substrate–air interface appears in the measured electric field at a temporal delay of 1.46\,ps. For the sample containing ITO nanocrystals, this reflection occurs 25\,fs earlier than for the pure substrate. Panels (a) and (b) illustrate the relative temporal separation of the excitation pulses, aligned such that their reflections at 1.46\,ps overlap in the measured electric fields.}
    \label{SI3}
\end{figure}

To quantify the modulation depth of the ITO nanocrytals, the excitation fluence was systematically varied between $10\,\frac{\mu\text{J}}{\text{cm}^2}$ and $8\,\frac{\text{mJ}}{\text{cm}^2}$ by controlling both the neutral-density attenuation and the beam focusing geometry. The additional dispersion introduced by the neutral density filters was compensated using a set of dispersive mirrors, ensuring Fourier transform-limited pulses at the sample. A second set of dispersive mirrors was used to compress and temporally image the excitation pulses at the field-sampling nonlinear medium. \textbf{Figure \ref{fig:3}a} presents the normalized Wigner–Ville distribution for both the substrate (reference) and the sample responses at fluences between $10\,\frac{\mu\text{J}}{\text{cm}^2}$ and $1.2\,\frac{\text{mJ}}{\text{cm}^2}$. As described above, the sample response was determined by subtracting the transmitted electric field through the substrate from that through the sample under identical fluence conditions. \textbf{Figure \ref{fig:3}b} and \textbf{Figure \ref{fig:3}c} show the temporal response of the sample and its corresponding Fourier transformation at the temporal window of -\,350\,fs to 400\,fs and at fluences of $10\,\frac{\mu\text{J}}{\text{cm}^2}$ and $1.2\,\frac{\text{mJ}}{\text{cm}^2}$, respectively. With increasing fluence, the response amplitude decreases, reflecting the gradual bleaching of the ITO nanocrystals and demonstrating their optical switching behavior. The associated increase in transmission arises from reduced attenuation, which may result from either a decrease in absorption or changes in the reflectivity of the sample. The transient sample response persists for up to 400\,fs, corresponding to a spectral bandwidth of 2.5\,THz. At longer time delays, the measured field also carries contributions from the absorption of atmospheric water molecules. The LSPR response remains detectable within this regime, imprinted on the water absorption features, albeit with significantly reduced relative amplitude (see Figure \ref{SI2} SI for details). For the substrate measurements, a reflection from the second surface is resolved at a time delay of 1.45 ps, consistent with the 150\,$\mu$m substrate thickness. For the ITO sample, this reflection is shifted forward by 25\,fs (see Figure \ref{SI3} SI for details). 

\begin{figure}[t]
    \centering
    \includegraphics[width=0.85\textwidth]{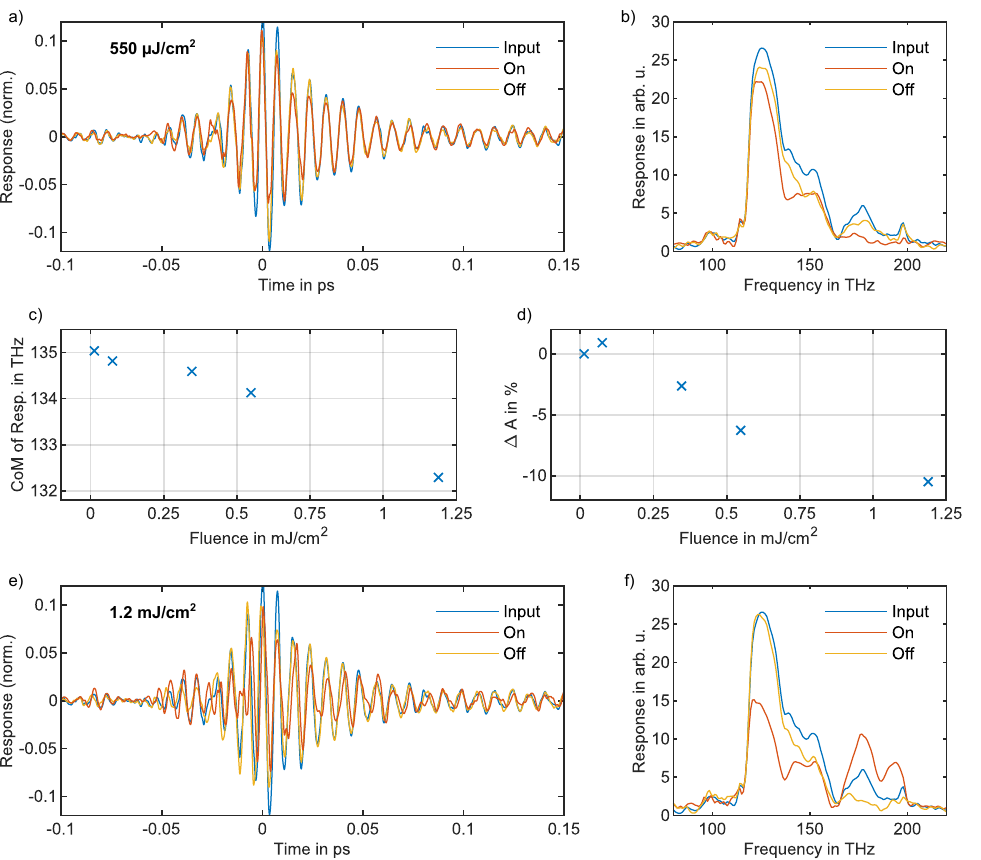}
\caption{a) Reversibility response of the optical switching at the fluence of $550\,\frac{\mu\text{J}}{\text{cm}^2}$. The ‘Input’ trace shows the response field at low fluence. ‘ON’ corresponds to the response field at $550\,\frac{\mu\text{J}}{\text{cm}^2}$, while ‘OFF’ represents the response field measured at low fluence after switching. b) The spectral response of switching at the fluence of $550\,\frac{\mu\text{J}}{\text{cm}^2}$. c) Weighted mean frequency of the response at various fluences. A red shift in the weighted mean frequency is observed as the fluence increases up to $1.2\,\frac{\text{mJ}}{\text{cm}^2}$. d) Modulation depth of the switch at various fluences. The integrated $\Delta$A is calculated by subtracting the integrated intensity of the ITO response at each fluence from that at $10\,\frac{\mu\text{J}}{\text{cm}^2}$, and normalizing by the integrated intensity of the reference. e) Reversibility response of the optical switching at the fluence of $1.2\,\frac{\text{mJ}}{\text{cm}^2}$. f) The spectral response of switching at the fluence of $1.2\,\frac{\text{mJ}}{\text{cm}^2}$.}
    \label{fig:4}
\end{figure}

\begin{figure}[ht]
    \centering
    \includegraphics{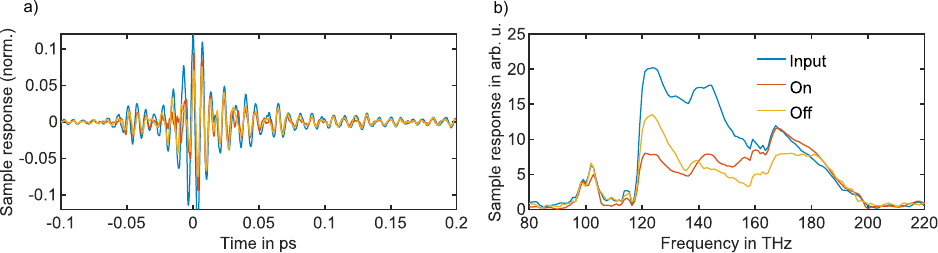}
    \caption{a) Irreversibility of the optical switching at the fluence of $3.3\,\frac{\text{mJ}}{\text{cm}^2}$. b) Spectral response of the switch at $3.3\,\frac{\text{mJ}}{\text{cm}^2}$.}
    \label{fig:5}
\end{figure}

\begin{figure}[htb]
    \centering
    \includegraphics{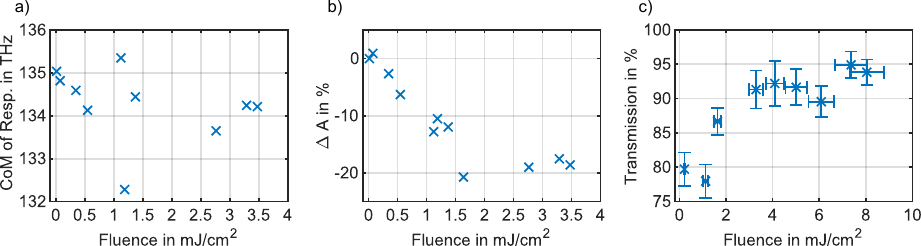}
    \caption{SI. a) Weighted mean frequency of the response. It can be seen that the red shift of the absorption disappears when the reversibility is compromised. b) Modulation depth (change in attenuation), obtained from the integrated spectral intensity of the measured field with and without ITO. The modulation depth remains constant at higher fluences. c) The measured transmitted power at various fluences by a thermal powermeter.}
    \label{SI4}
\end{figure}

\begin{figure}[htb]
    \centering
    \includegraphics{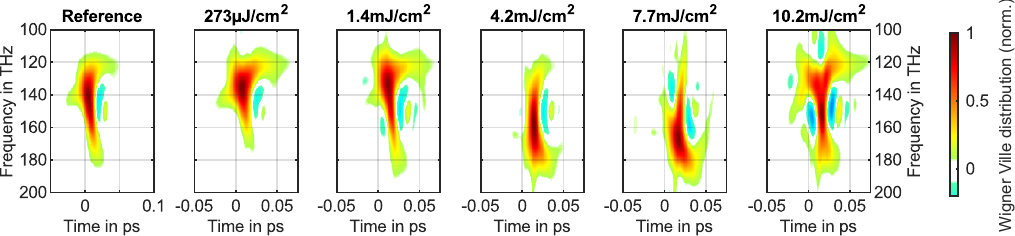}
    \caption{SI. Normalized Wigner–Ville distribution (WVD) of the substrate and sample responses at fluences between $215\,\frac{\mu\text{J}}{\text{cm}^2}$ and $8\frac{\text{mJ}}{\text{cm}^2}$. At low fluence, the response spectrum is dominated by the LSPR, whereas at higher fluences, the LSPR feature progressively bleaches. At $6\,\frac{\text{mJ}}{\text{cm}^2}$ the WVD exhibits a pronounced blue shift. At $8\,\frac{\text{mJ}}{\text{cm}^2}$ the WVD closely resembles that of the reference, indicating a spectrally uniform intensity loss.}
    \label{SI5}
\end{figure}

The practical applicability of any switch requires reversible behavior following excitation. Accordingly, a series of measurements is carried out to assess whether the sample reverts to its initial state after exposure to high fluence under 1\,MHz duty-cycle. \textbf{Figure \ref{fig:4}a} and \textbf{Figure \ref{fig:4}b} shows the reversibility of the sample at $550\,\frac{\mu\text{J}}{\text{cm}^2}$ fluence, corresponding to modulation depth of 6\%. At this fluence, the qualitative attenuation behavior remains largely unchanged, and the switch fully recovers to its initial state once deactivated. With increasing fluence, the weighted mean frequency of the spectral response exhibits a red-shift, and the modulation depth increases, reaching 10\% at $1.2\,\frac{\text{mJ}}{\text{cm}^2}$ (see \textbf{Figure \ref{fig:4}c} and \textbf{Figure \ref{fig:4}d}). At a fluence of $1.2\,\frac{\text{mJ}}{\text{cm}^2}$, the switch still exhibits reversibility. Under these conditions, the overall transparency of the sample increases, and the spectral bandwidth of the response broadens to match that of the excitation pulse, suggesting modifications in reflectivity at the sample interfaces. However, a slight alteration of the spectral distribution is observed after the switch is turned off (see \textbf{Figure \ref{fig:4}e} and \textbf{Figure \ref{fig:4}f}). Higher fluences lead to larger modulation depths and broaden the response bandwidth. For fluences above $1.2\,\frac{\text{mJ}}{\text{cm}^2}$, a qualitative blue shift of the response emerges, indicating enhanced transparency across the full spectral range of the excitation pulse (see Figure \ref{SI4} and Figure \ref{SI5} in SI). However, beyond a fluence of $3.3\,\frac{\text{mJ}}{\text{cm}^2}$, the switch becomes fully irreversible, as shown in \textbf{Figure \ref{fig:5}a} and \textbf{Figure \ref{fig:5}b}. 

\begin{figure}[htb]
    \centering
    \includegraphics{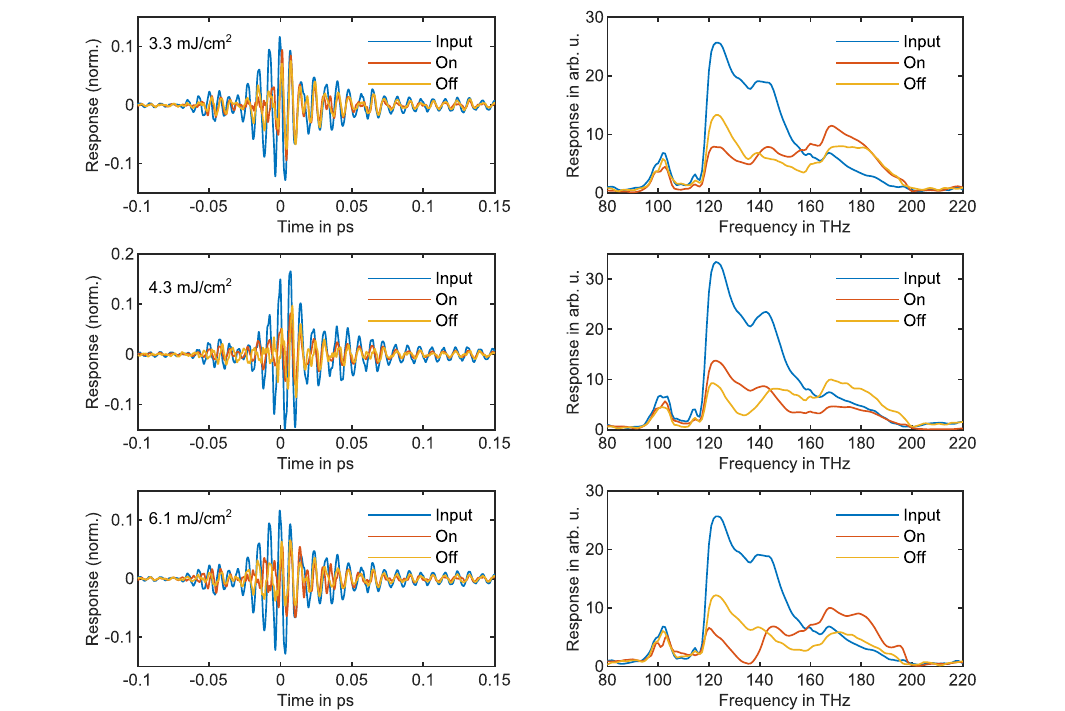}
    \caption{SI. Nonlinearity and reproducibility of the optical switching of the ITO nanocrystals at fluences beyond 3.3\,$\frac{\text{mJ}}{\text{cm}^2}$. It is seen that the switch becomes fully irreversible at these fluences.}
    \label{SI6}
\end{figure}

The electric field response of the sample captures sub-cycle details of its nonlinear dynamics across different fluences. Analysis of the transient electric field in Figure \ref{fig:4}a, Figure \ref{fig:4}e, and Figure \ref{fig:5}a indicates that the sample response in the first cycle of the excitation pulse progressively weakens with increasing fluence (see Figure \ref{SI6} SI for more details). Upon pump excitation in plasmonic systems, a sequence of processes unfolds on distinct timescales, including electron dephasing, electron–electron scattering, electron–phonon coupling, and lattice heat dissipation. In nanocrystals, the faster carrier dynamics and the observed 400 fs recovery time can be attributed to abundant surface trapping states, which provide additional decay pathways for photogenerated hot electrons compared to ITO films and nanorods \cite{guo2016ultrafast,hartland2011optical}. This ultrafast response can be divided into three stages: an initial photoinduced bleaching phase dominated by electron dephasing, a subsequent phase governed by electron–electron scattering, and a final recovery phase driven by electron–phonon coupling and lattice heat dissipation \cite{Guo2017SolutionProcessed}.

The irreversibility of the switch can be attributed to the damage and deformation of the nanocrystals or the substrate caused by the high peak power of the excitation pulses or their high duty cycle. The corresponding peak intensity for the excitation pulses at $1.2\,\frac{\text{mJ}}{\text{cm}^2}$ and $3.3\,\frac{\text{mJ}}{\text{cm}^2}$ is $664\,\frac{\text{GW}}{\text{cm}^2}$ and $1.8\,\frac{\text{TW}}{\text{cm}^2}$, respectively. The damage threshold of ITO films excited by nanosecond pulses at repetition rates from a few hertz to tens of kilohertz is on the order of $300\,\frac{\mu\text{J}}{\text{cm}^2}$ \cite{yoo2017optical,peng2019high}. Higher thresholds corresponding to $1\,\frac{\text{TW}}{\text{cm}^2}$ were reported for excitation pulses as short as 35\,fs at kilohertz duty cycles \cite{Huang2022BroadbandSaturableAbsorptionITO}. It should be noticed that the fieldoscopy measurements reveal that switching at these intensities tends to become partially or fully irreversible. 

\section{Conclusion}

An effective optical switch must operate at high duty cycles while maintaining a substantial modulation depth and full reversibility. In this work, we have investigated the extent to which ITO nanocrystals excited within their LSPR spectral range fulfill these requirements. We employed the novel technique of fieldoscopy, which offers high detection sensitivity, to resolve the electric field response of ITO nanocrystals excited by two-cycle femtosecond pulses within their LSPR spectral range at megahertz repetition rates. Our study reveals that the LSPRs of the ITO nanocrystals can be excited at high peak intensities reaching $1.8\,\frac{\text{TW}}{\text{cm}^2}$. However, the reversibility of the switching degrades for peak intensities beyond $664\,\frac{\text{GW}}{\text{cm}^2}$. This is crucial information for the practicality of employing optical switches. 

The fluence-dependent bleaching and redshift of the attenuated field is reversible up to $1.2\,\frac{\text{mJ}}{\text{cm}^2}$ and becomes fully irreversible for fluences beyond $3.3\,\frac{\text{mJ}}{\text{cm}^2}$. Employing two-cycle excitation pulses in combination with fieldoscopy enables access to the sub-cycle dynamics of the excitation process. It is observed that the response of the sample begins rising during the first optical cycle of the pulse, reaching its maximum in the second cycle. However, increasing the excitation fluence reduces the relative contribution from the first cycle, and the response is dominated by the second optical cycle of the excitation pulses. The demonstrated reversible transmission modulations at a megahertz duty cycle are particularly significant, as optical modulators with high duty-cycle operation and fast recovery times are increasingly sought after in light of advances in Yb:YAG laser systems and SWIR pulse generation \cite{fattahi2014third, alismail2020multi, alismail2017carrier, wang2017cross, buberl2016self, li20240, saraceno2019amazing}. The colloidal chemistry approach for preparing the nanocrystals offers a scalable and versatile route for printable photonics and optoelectronics. These results open a pathway toward ultrafast, robust, and broadband optical modulators based on engineerable conducting metal oxides, with applications extending well beyond the present work \cite{rebecchi_transparent_2023, guizzardi_near-infrared_2022, diroll_large_2016, noginov2011transparent, wu2012metamaterial,wu2012fano,tittl2015switchable,kohoutek2012integrated}.

\section{Method} 
\begin{figure}[t]
    \centering
    \includegraphics{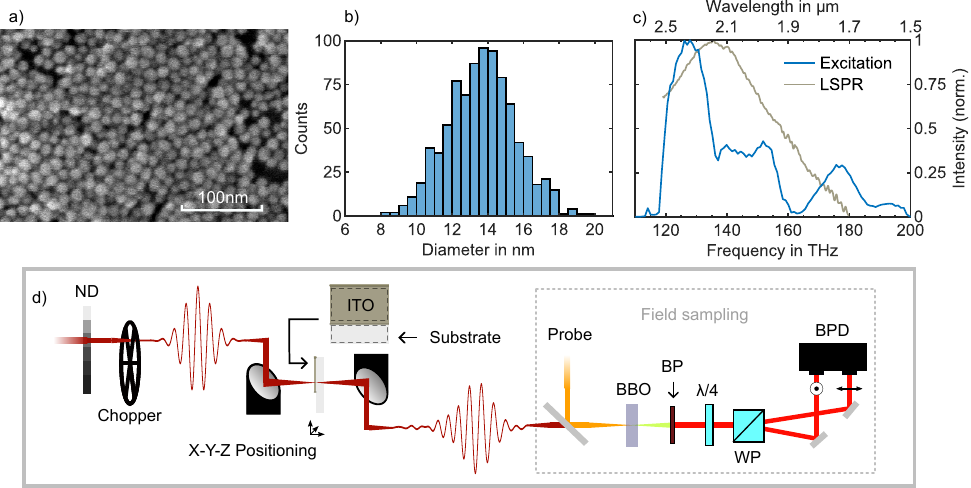}
    \caption{a) Top-view scanning electron microscopy (SEM) micrograph of ITO nanocrystal thin film. b) Statistical evaluation of the nanocrystals' diameter. In total, 871 crystals over an area of 470\,nm$\times$650\,nm  are analyzed, resulting in a mean nanocrystal diameter of 14\,nm. c) Spectrum of the 10.7\,fs excitation pulses (blue curve). The LSPR spectrum of the nanocrystals (gray curve), measured by Fourier transform infrared spectroscopy (FTIR). d) Schematic of the experiment. The sample consists of a borosilicate substrate with two areas of ``ITO-coated'' and ``uncoated''. The uncoated section serves as a reference for the measurement. The sample is positioned at the excitation beam focus, while the beam is modulated at 10\,kHz by a mechanical chopper placed before the sample in the optical path. The fluence is varied from  $10\,\frac{\mu\text{J}}{\text{cm}^2}$ to $8\,\frac{\text{mJ}}{\text{cm}^2}$ by adjusting both a variable neutral-density (ND) filter and the beam focus size. The transient field is resolved by employing fieldoscopy. ND: variable neutral density filter; BBO: beta barium borate; BP: bandpass filter; $\lambda/4$: quarter wave plate; WP: Wollaston prism; BPD: balanced photo detector.}
    \label{fig:1}
\end{figure}

The front end is based on a Yb:KGW laser (CARBIDE from Light Conversion) providing 20\,µJ pulses at 1\,MHz repetition rates, centered at 1030\,nm, and at a duration of 255\,fs. The output pulses of the laser were compressed in two stages of gas-filled hollow core photonic crystal fibres to 4.6\,fs. 4\% of the power was directly sent to the sampling crystal as a probe pulse. The remaining power of 12\,µJ was sent to a BiBO crystal to generate the passively CEP stable excitation pulses spanning from 1500\,nm to 2500\,nm (120\,THz to 200\,THz) via intrapulse difference frequency generation (Figure \ref{fig:1}c) \cite{srivastava_180_2024, amotchkina2016broadband}. 

 For field resolved detection, the probe pulse and the excitation pulse were overlapped in a 20\,µm-thick type\,II BBO crystal, for sum frequency generation (SFG). The spectrum of the SFG partially overlaps with the high-frequency tail of the probe pulses. The overlapped region between 600\,nm to 700\,nm was isolated with a spectral filter. The cross-polarised filtered SFG and probe pulses were converted to circular polarisation in a $\lambda/4$ plate and sent to a Wollaston prism. A balanced detector was used to detect the orthogonal polarisation components. The excitation beam was periodically modulated with a mechanical chopper at 10\,kHz and the signal of the balanced detector was detected with a lock-in amplifier (Figure \ref{fig:1}d). A commercial interferometer (SmarAct GmbH, PICOSCALE) was used to measure the temporal offset of the probe and the excitation pulse. The electric field of the excitation pulses was resolved by temporally scanning the probe pulses, allowing for a detection dynamic range of 110\,dB, and detection sensitivity down to atto-Joule with a temporal jitter of 90\,as \cite{srivastava_near-petahertz_2024}.

The sample was positioned at the focus. The excitation pulses were compressed to a 10.7\,fs duration at the sample and at the field-sampling nonlinear crystals, by using ten bounces on dispersive mirrors before the sample, and 2 additional bounces after the sample. This way, it was ensured that the excitation pulse is optimally compressed in both of these points of interest. The intensity of the excitation pulse was controlled with a variable ND filter (Thorlabs NDL-25S-4) and focusing geometry. The sample is positioned in the focus of a 2\,inch focal length parabolic mirror, with a spot size of 30\,$\mu$m (full width at half maximum (FWHM)) for the regime of low fluence, and a 1\,inch parabolic mirror with a spot size of 21.7\,$\mu$m (FWHM) for the regime of high fluence.

The ITO samples employed in this study were prepared by dip-coating borosilicate glass substrates (150\,µm thickness, 20\,mm$\times$20\,mm lateral size) into 20\,mL of a suitably diluted, colloidally stable ITO dispersion. The dispersion was synthesized through a continuous growth method and dispersed in octane \cite{rebecchi2025scalable}. Optimal deposition parameters, resulting in approximately one monolayer per dip, were identified as follows: a concentration of 1\,mg/mL, a single dipping step with an immersion speed of 100\,mm/min, followed by a withdrawal step including a waiting time of 10\,s. The coated substrates were subsequently annealed at 100\,°C for 5\,min (see Figure \ref{SI1} SI). The annealing temperature affects the amount of residual ligand (oleic acid), which is used to fabricate a stable colloidal solution of nanocrystals. At an annealing temperature of 100\,°C, the ligands are expected to remain, which affects the dielectric environment of the nanocrystals as well as the geometric structure of the film and consequently the LSPR frequency \cite{garcia_dynamically_2011}. After the coating was applied, the film on one side of the cover glass was removed. Figure \ref{fig:1}a shows a scanning electron microscopy (SEM) image of the ITO nanocrystals with a mean size of $14\pm2\,$nm (Figure \ref{fig:1}b). The sample features a section of exposed substrate, which has not been dipped into the colloidal solution. This area served as a reference point. To select between ITO and the substrate, the sample was moved via a manual three-way translation stage in the focus of the excitation pulse, such that the light is either focused on the ITO or the substrate.

\medskip
\textbf{Acknowledgements} \par 
HF acknowledges funding from the ERC Consolidator Grant ID: 101125670. IK acknowledges the support from the ERC and EIC Grants ID: 850875 and 101186701. AH acknowledges the support from Max Kieker, Pragya Sharma, and Eduard Butzen.

\medskip

%
\bibliographystyle{MSP}
\bibliography{Sources}

\newpage

\newpage
\section*{Supplementary Information}

\end{document}